# Forecasting Electricity Smart Meter Data Using Conditional Kernel Density Estimation

Siddharth Arora[†]and James W. Taylor[*]

Saïd Business School,
University of Oxford, Park End Street, Oxford, OX1 1HP, U.K.

[†] Siddharth Arora
Tel: +44 (0)1865 288800
Email: Siddharth.Arora@sbs.ox.ac.uk

[*]James W. Taylor
Tel: +44 (0)1865 288800
Email: James.Taylor@sbs.ox.ac.uk

# Forecasting Electricity Smart Meter Data Using Conditional Kernel Density Estimation

*Abstract* – The recent advent of smart meters has led to large micro-level datasets. For the first time, the electricity consumption at individual sites is available on a near real-time basis. Efficient management of energy resources, electric utilities, and transmission grids, can be greatly facilitated by harnessing the potential of this data. The aim of this study is to generate probability density estimates for consumption recorded by individual smart meters. Such estimates can assist decision making by helping consumers identify and minimize their excess electricity usage, especially during peak times. For suppliers, these estimates can be used to devise innovative time-of-use pricing strategies aimed at their target consumers. We consider methods based on conditional kernel density (CKD) estimation with the incorporation of a decay parameter. The methods capture the seasonality in consumption, and enable a nonparametric estimation of its conditional density. Using eight months of half-hourly data for one thousand meters, we evaluate point and density forecasts, for lead times ranging from one half-hour up to a week ahead. We find that the kernel-based methods outperform a simple benchmark method that does not account for seasonality, and compare well with an exponential smoothing method that we use as a sophisticated benchmark. To gauge the financial impact, we use density estimates of consumption to derive prediction intervals of electricity cost for different time-of-use tariffs. We show that a simple strategy of switching between different tariffs, based on a comparison of cost densities, delivers significant cost savings for the great majority of consumers.

# 1. Introduction

A smart meter is an electronic device that measures electricity consumption at the installed facility, and transmits this information to the consumer and the energy supplier/operator on a near real-time basis. It is anticipated that based on smart meter information, significant energy and financial savings can be achieved through detailed consumer feedback and tariffs designed for facilitating energy savings (CER 2011a; 2011b). The large scale installations of smart meters will generate massive amounts of data, offering unique insight into the consumption behaviour of different consumers. Over the coming years, smart meters are scheduled to replace the existing electronic meters. It is estimated that by 2019, approximately 60 million meters will be installed and operable in the United States (Edison Foundation Report, 2012). In the European Union, all member states must have smart meters installed for at least 80% of consumers by 2020, with full deployment by 2022 (Directive 2009/72/EC). It has been estimated that the cost of investment in smart electricity grids in the European Union will be around €51 billion (Faruqui *et al*., 2010).

Unlike conventional meters, smart meters provide site-specific information regarding electricity consumption throughout the day. This information can potentially change the landscape of energy markets, by allowing suppliers to make highly data-dependent decisions to develop innovative dynamic pricing strategies for their target consumers. Smart meters, along with different time-of-use (TOU) tariffs, can help consumers shift their consumption away from peak hours, which can result in significant savings (Cosmo *et al*., 2012). With the liberalization of electricity markets, market participants rely on accurate forecasts to make informed energy transactions (Woo *et al*., 2006). Also, smart meters can assist electricity scheduling, thereby facilitating safe and efficient operation of the power system.

In a recent trial involving electricity smart meter installations in Ireland, the deployment of TOU tariffs and information stimuli, such as bi-monthly billing and an in-home display device, resulted in an overall reduction in electricity usage by 3.2% and peak usage by 11.3% (CER 2011a; 2011b). Given the potential for smart meters in enabling the efficient use of energy, and its financial implications for energy markets, it is imperative to develop accurate methods for modelling electricity smart meter data. This is the focus of this paper.

Electricity consumption data from individual smart meters exhibits seasonality, comprising both intraday and intraweek cycles, and in comparison with total national demand, consumption recorded from individual smart meters exhibits much higher variability. It, therefore, seems appropriate to provide a forecast of the probability density function for consumption recorded by a smart meter, rather than just producing a point forecast. Focusing on the density forecasts of electricity consumption from smart meters can help: a) ensure that the risk associated with complex decision making based on such forecasts is adequately assessed, and, b) generate accurate consumption estimates at varying aggregation levels, potentially resulting in improved demand side management. The literature on modelling electricity smart meter data is small. There are some recent papers on short-term point forecasting of smart meter data (Ghofrani *et al.*, 2011; Ding *et al.*, 2011), and forecast error measures (Haben *et al.*, 2013). However, we are not aware of existing studies on modelling the density of electricity consumption data from individual smart meters.

The non-Gaussian and highly variable nature of individual smart meter electricity consumption data motivates nonparametric probability density estimation methods. In this study, we propose methods based on kernel density (KD) and conditional kernel density (CKD) estimation (see, for example, Rosenblatt, 1969; Hyndman *et al.*, 1996). The conditioning in our CKD implementations aims to capture seasonality. Although this seasonality is usually far less clear than the seasonality in a series of the total electricity demand for a country, both types of data tend to exhibit intraday and intraweek seasonal cycles. This prompts us to consider forms of KD and CKD that are inspired by the structure of models presented in the literature for total national consumption, namely the exponential smoothing and autoregressive moving average (ARMA) models presented by Taylor (2003, 2010) and Gould *et al.* (2008). The CKD method involves kernel weighting over the conditioning variable. We consider different implementations of the CKD estimator, where we condition consumption on the period of week, period of day, and lagged consumption. We chose methods based on KD and CKD estimation for modelling smart meter data, because these methods: a) model the full density function, b) can accommodate seasonality in the time series, and, c) make no distributional assumption for the shape of the density, allowing the estimated density to be, for example, multi-modal, fat-tailed or skewed. The incorporation of a decay factor within the CKD estimation helps model temporal evolution in the relationship between consumption and the conditioning variables.

This paper employs eight months of half-hourly data, recorded from one thousand smart meters, to evaluate the KD and CKD methods, in terms of accuracy of their density, quantile, and point forecasts, for lead times ranging from one half-hour up to one week ahead.

Although weather variables are often used in energy modelling (Regnier, 2008), the KD and CKD methods that we propose use only the historical consumption observations. We felt that we could not assume the availability and affordability of weather predictions for a location reasonably close to each smart meter. Furthermore, the use of weather data in a large-scale online prediction system raises issues of robustness (Bunn, 1982).

Recent advances in smart metering technology may pave the way for easy switching between suppliers, and between different payment schemes from the same supplier (Darby, 2012). In this study, we use our density estimates of electricity consumption to derive prediction intervals for electricity cost, for different TOU tariffs. We compare different costs that would potentially be incurred in the future, for each available tariff, and select the tariff that would result in the greatest cost savings. In a case study of a warehouse environment, Sanders and Graman (2009) emphasize the importance of evaluating the impact of forecast errors on organizational cost.

In Section 2, we describe the smart meter data. Section 3 presents different KD and CKD methods, along with an exponential smoothing benchmark method. Empirical results regarding forecast accuracy are provided in Section 4. Section 5 derives prediction intervals for electricity cost. Section 6 summarizes and concludes the paper.

## 2. Smart meter data

We used eight months of half-hourly smart meter data for electricity consumption from 2 January to 31 August 2010. We used the first seven months, comprising 10128 observations (*in-sample*), for optimizing method parameters, while the final month, constituting 1488 observations (*post-sample*), was used to evaluate forecast accuracy. The final month of the in-sample period was used for cross-validation. Using a moving window of six months, we generated a sequence of density forecasts, for lead-times ranging from one half-hour up to one week ahead, by using as forecast origin each midnight in the post-sample data. The data was recorded for 800 residential consumers and 200 small to medium-sized enterprises (SMEs). The data was obtained from the Commission for Energy Regulation (CER) based in Ireland (CER 2011a; 2011b). Figs. 1 and 2 show consumption for a residential consumer and an SME, respectively. The figures show that consumption for a residential consumer is rather volatile, and consumption for the SME displays a prominent seasonal pattern.

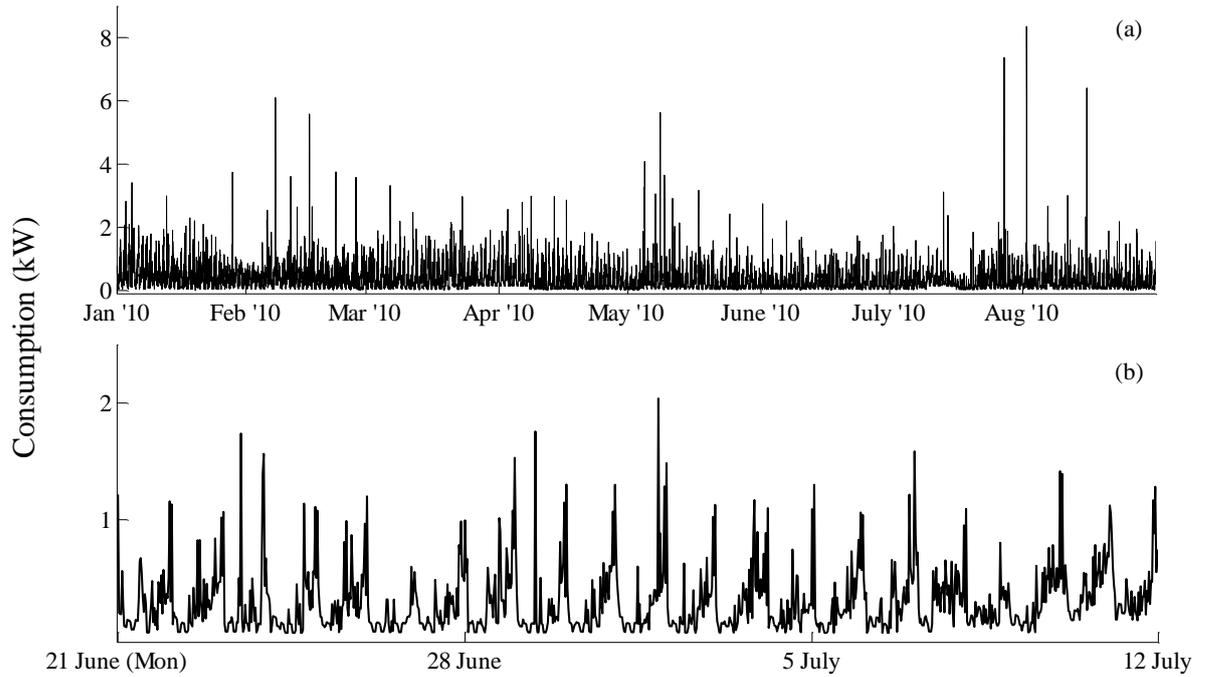

**Fig. 1.** Consumption for a residential consumer for: a) the full eight months, and, b) a typical three-week period.

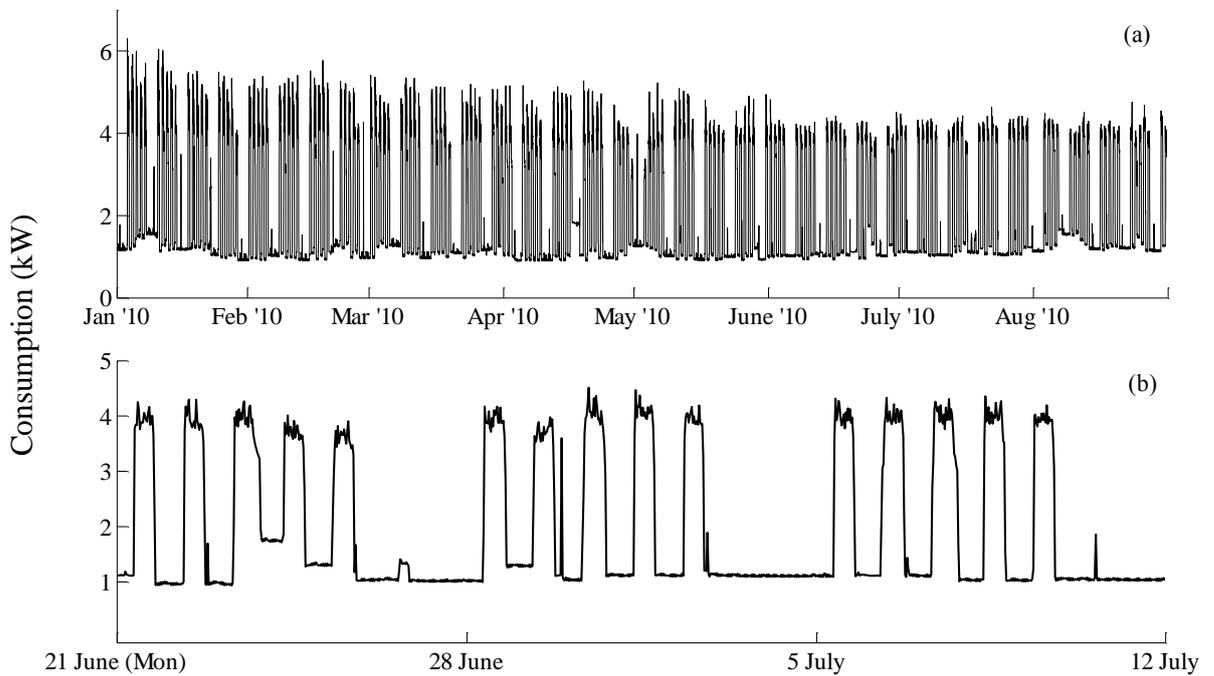

**Fig. 2.** Consumption for an SME for: a) the full eight months, and, b) a typical three-week period.

Fig. 3 presents the average daily (intraday) cycle for the same residential consumer and SME considered in Figs. 1 and 2. We calculated the averages shown in Fig. 3 using only the in-sample data. For the residential consumer, consumption does not differ noticeably across the different days of the week, whereas for the SME, consumption is significantly higher on weekdays than on weekends. Furthermore, consumption for the residential consumer is high during the evenings, while for the SME, consumption is high only during the typical working hours of weekdays. Figs. 1 to 3 indicate that the data exhibits, to varying degrees, repeating intraweek seasonal cycles, and also, at least for the weekdays, repeating intraday cycles. The density estimation methods proposed in this study aim to capture this double seasonality. We represent the lengths of the intraday and intraweek cycles by $s_1$ and $s_2$, respectively. Since the smart meter data is recorded half-hourly, we have $s_1 = 48$ and $s_2 = 336$.

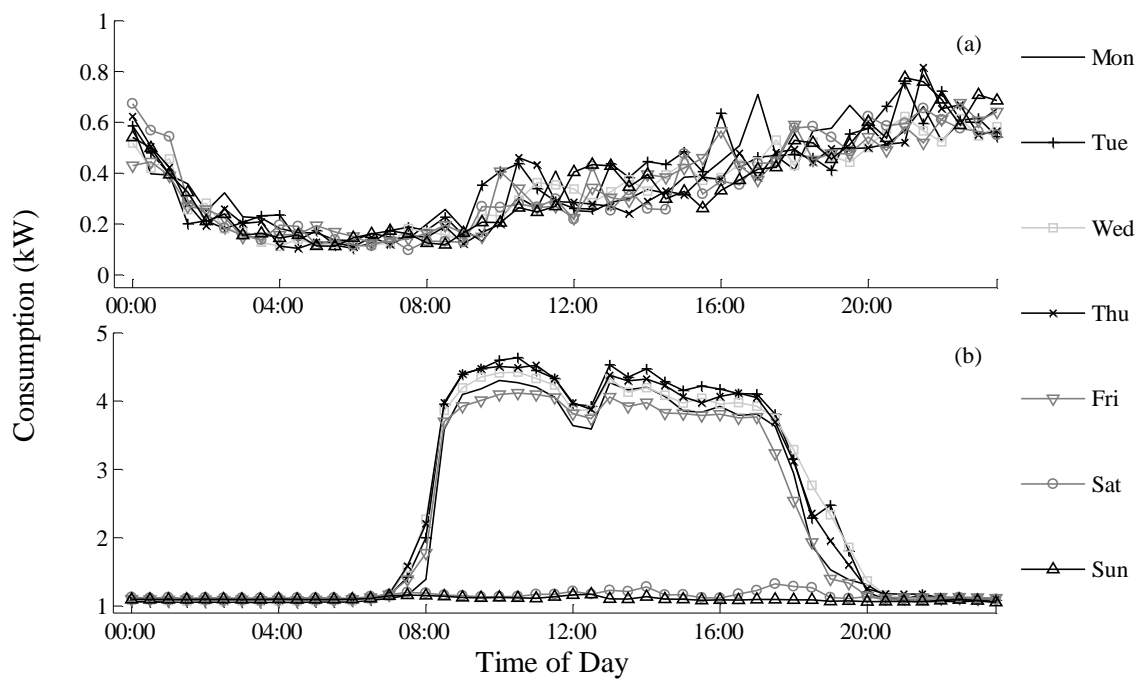

**Fig. 3.** Average intraday profile for each day of the week for: (a) a residential consumer, and, (b) an SME.

Prior to modelling, for each consumer, consumption observations were divided by their maximum value, so that the observations fell between 0 and 1. This standardization allows, for each method, averaging of accuracy metrics across consumers.

We identified seven special days, e.g. public holidays, in the data. Since most SMEs do not operate on public holidays, the consumption on such days tended to be considerably lower than if it had been a normal working day. For special days lying within the estimation sample (for both residential consumers and SMEs), we smoothed out the anomalous consumption observations, by replacing them with recent historical observations belonging to a normal working day. Just one special day fell in the evaluation sample, namely Monday 2 August, which was a public holiday in Ireland. For this day, we adjusted the forecasting approach for each of the various methods that we implemented. For each consumer, we first compared the actual consumption for the last special day observed in the estimation sample, namely the public holiday on Monday 7 June, with the consumption observed on: a) the corresponding day of the week from the previous week, and b) the most recent Sunday. If the consumption observed on Monday 7 June was more similar to a normal working day than a Sunday (as quantified using the mean absolute difference), we produced forecasts for 2 August as if it was a normal working day. Otherwise, we treated 2 August as a Sunday. Future work could investigate alternative approaches to accommodating such holiday effects, such as the rule-based approach of Arora and Taylor (2013), developed for national demand data.

Electricity retailers will have many thousands, and quite possibly millions, of smart metered customers. It is unlikely to be practical to estimate separately the parameters of a density estimation method for the time series of consumption recorded by each meter. Estimation of method parameters based on just a sample of the series is, therefore, required. It seems natural to surmise that method parameters will be more similar for consumers with similar consumption characteristics, than for those with very different consumption patterns. This motivates the categorization of the series, according to consumption characteristics, followed by the estimation of method parameters for a sample of the consumers from each category. In our empirical analysis for this paper, we used the same consumption data as used by the CER, and so we employed their categorization scheme (see CER, 2011a). This scheme was based on a principle component analysis of customer characteristics, namely, demographic, household and socio-economic parameters. This categorization dictated the allocation of the consumers to different tariffs and stimuli.

The CER divided consumers into different control and treatment groups. The bills for the control group were based on their normal tariff, whereas consumers within the treatment groups were subjected to different TOU tariffs. Specifically, for the residential consumers in the treatment group, there were five different tariff types, corresponding to different TOU prices. For the SMEs, there were two different tariff types. The five residential TOU tariffs are presented in Table 1. The day period was defined as 8am-5pm and 7pm-11pm on weekdays, and 5pm to 7pm on weekends and public holidays. The peak period was 5pm-7pm on weekdays (excluding bank holidays) and the night period was 11pm-8am. For the weekend tariff, night rates were applicable for all periods on weekends, while on weekdays (excluding bank holidays), different day, peak, and night rates were applicable

**Table 1**   Time-of-use tariffs (cents per kWh) for residential consumers.

| Tariff | Day | Peak | Night |
|---|---|---|---|
| A | 14 | 20 | 12 |
| B | 13.5 | 26 | 11 |
| C | 13 | 32 | 10 |
| D | 12.5 | 38 | 9 |
| Weekend | 14 | 38 | 10 |

In addition to being allocated a specific tariff, consumers within the treatment group were provided with different stimuli, i.e., information regarding their electricity usage. There were four different stimuli for residential consumers (see CER, 2011a), namely, a) bi-monthly billing; b) monthly billing; c) bi-monthly billing with an in-home device that provides real-time consumption information; and, d) bi-monthly billing with an overall usage reduction incentive. Consumers within the control group and the weekend TOU tariff group did not receive any stimulus. We categorized the consumers according to the different tariff and stimulus types. This implies a total of 18 different categories for residential consumers. The SMEs were allocated two tariffs; however, for each SME, we only had information regarding the stimulus. Hence, for the purpose of model estimation, we group SMEs based only on stimulus in this study. The SMEs were provided with four stimulus plans, a) monthly detailed bill; b) bi-monthly detailed billing with an in-home device; c) bi-monthly detailed bill; and d) bi-monthly detailed billing with consumption and cost information made available via a web account. Taking into account the control group, the SMEs were classified into 5 categories.

To reduce the computational time, and allow scalability of the proposed modelling approach, we estimated method parameters for up to 10% of the consumers within each category. The median of the optimized values for each parameter was then used for all consumers in the corresponding category. The same process was then repeated for all different categories separately, in order to generate post-sample forecasts for all one thousand consumers. Fildes *et al*. (1998) adopted a similar approach for parameter optimization, and reported that using the most commonly occurring parameter values leads to better forecasting results, compared to the case when parameters are either chosen arbitrarily, optimized once, or optimized at each forecast origin.

## 3. Forecasting methods

In this section, we first briefly describe standard kernel density estimation, and conditional kernel density estimation. There are two notable challenges with the implementation of these methods: the selection of the kernel bandwidths, and the presence of boundary effects. In Section 3.2, we discuss the strategy adopted in this paper to overcome these two challenges, and in Sections 3.3-3.9, we present the different kernel-based methods that we implemented for our smart meter data. In Section 3.10, we summarise the parameter estimates from these methods. In Section 3.11, we present an exponential smoothing benchmark method.

### 3.1 Unconditional and conditional kernel density estimation

Nonparametric density estimation has received considerable attention due to its widespread application in statistical data analysis (see, for example, Hastie *et al*., 2009). Kernel density (KD) estimation allows the nonparametric estimation of a density $f$ for a set of observations $\{Y_1, Y_2, ..., Y_n\}$. In this paper, the observations are for the electricity consumption of an individual consumer. Recall that, as we explained in the previous section, for each consumer, we standardized the consumption values to lie between 0 and 1. The kernel density estimator can be represented as:

$$\hat{f}(y) = \sum_{t=1}^{n} K_{h_y}(Y_t - y), \quad (3.1)$$

where $\hat{f}(y)$ denotes the local density estimate, at point $y$; $K_{h_y}(\bullet) = K(\bullet/h_y)/h_y$ denotes the kernel function; and $h_y$ is the bandwidth of the kernel function. A kernel is essentially a weighting function centred at a given point, whereby the rate of weight decay is controlled by the kernel bandwidth.

We used a Gaussian kernel function for all density estimation methods employed in this paper. To generate a full density, we repeated the density estimation for a range of different values of $y$. Given that consumption observations for most of the consumers are concentrated towards zero, we divided the range of $y$ into one hundred non-uniform increments. Specifically, for each consumer, we determined the first ninety values of $y$ by dividing the range between 0 and the 90[th] percentile of the observations into ninety uniform increments. The remaining values of $y$ were determined by dividing the interval between the 90[th] percentile and 1, into ten uniform increments.

In this study, in addition to KD estimation, we also investigated a *conditional* kernel density (CKD) estimator, which involves kernelling in both the $x$ and $y$-directions (Rosenblatt, 1969). The standard KD estimation involves only one kernel in the $y$-direction, and we refer to this as *unconditional* KD estimation. The CKD estimator produces an estimate of the conditional density $f(y|x)$ of a variable $y$, conditional on the variable $x$. The essential idea is that, for a given *x*, the density function at the value *y* is constructed by applying kernel density estimation to the set of observations $\{Y_1, Y_2, ..., Y_n\}$, with each $Y_t$ value weighted in accordance with the closeness of the corresponding $X_t$ to the conditioning value $x$. In contrast to the classical linear regression based approach, which focuses on modelling the conditional expectation $E(y|x)$ in a parametric framework, CKD allows a nonlinear and nonparametric modelling of the whole density. With CKD, the density of a dependent variable $Y_t$, denoted by $f(y|x)$, conditional on $X_t = x$, is represented as:

$$\hat{f}(y|x) = \frac{\sum_{t=1}^{n} K_{h_x}(X_t - x) K_{h_y}(Y_t - y)}{\sum_{t=1}^{n} K_{h_x}(X_t - x)}, \quad (3.2)$$

where $h_x$ is the additional bandwidth in the $x$-direction.

The literature is rather limited on applications of the CKD estimator in the context of time series forecasting. Hyndman *et al*. (1996) used the CKD estimator to model daily temperature, by conditioning current temperature on lagged temperature observations and a seasonal variable. Ruan (2010) employed the CKD estimator for real-time prediction of the respiratory response. Bessa *et al*. (2012) proposed a recursive time-adaptive CKD estimator, which adapts with the arrival of new observations without re-computing the entire density from scratch. Jeon and Taylor (2012) used the CKD estimator to predict the density of wind power conditional on wind velocity densities.

*3.2    Kernel bandwidth selection and boundary correction*

It is crucial to estimate the kernel bandwidth appropriately, as a large bandwidth leads to over-smoothing of the main underlying features of the time series, whereas a small bandwidth leads to under-smoothing, such that the estimated density is too rough. There are two approaches proposed in the literature to estimate the kernel bandwidths: rule-based and data-based. Rule-based approaches select the bandwidths using a pre-determined rule, which is optimal under certain assumptions for the type of density (see, for example, Bashtannyk and Hyndman, 2001). Data-based approaches make no strong prior density assumptions, and estimate the bandwidths using the available data (Fan and Yim, 2004). In comparison with rule-based approaches, data-based approaches require more computational time. Fan and Yim (2004) give support to the use of the data-based approach of cross-validation. We used this approach in this paper. Specifically, we estimated the bandwidths by minimizing the one-step ahead prediction error of the density estimates, using the one month cross-validation hold-out sample. As discussed further in Section 4, we quantify density forecast accuracy using the continuous ranked probability score (CRPS) (see Gneiting *et al.*, 2007).

A potentially serious problem with nonparametric density estimation is the presence of boundary effects. Boundary effects correspond to the inclusion of only one-sided information at the boundaries, which results in the biased estimation of the final density. Since we transformed consumption to be between 0 and 1, these are the lower and upper boundaries in our application. For values of *y* that are close to these boundaries, a symmetric kernel will erroneously allocate non-zero weights beyond the boundaries, resulting in an estimated density that extends outside the range of 0 to 1.

In this study, we adopted the boundary correction approach proposed by Dai and Sperlich (2010). This approach involves the adjustment of the bandwidth for density estimation near the boundaries:

$$h_y = \begin{cases} \max(y, \varepsilon) & if\ 0 \leq y < h_1, \\ \max(1 - y, \varepsilon) & if\ (1 - h_1) < y \leq 1, \\ h_1 & otherwise \end{cases} \quad (3.3)$$

where $h_1$ denotes the default bandwidth. The term $\varepsilon$ was included for numerical reasons to avoid the bandwidth becoming zero. We specified $\varepsilon = 0.001$. The rationale of this method is to reduce the bandwidths in the boundary area, in order to prevent the kernel from allocating non-zero weights outside the upper and lower boundary limits. The extent of bandwidth reduction is determined by the closeness of *y* to the boundary limits. We incorporated boundary correction during parameter estimation and post-sample evaluation, for all kernel based methods considered in this study.

*3.3    Unconditional kernel density estimation (KD-U)*

With this benchmark method, we applied the KD estimation to a moving window of the most recent observations. This method, which we refer to as KD-U, is represented as:

$$\hat{f}(y) = \sum_{t=m}^{n} K_{h_y}(Y_t - y), \quad (3.4)$$

where $m = n - l + 1$, such that $l$ denotes the length of the moving window, $n$ corresponds to the forecast origin. In this study, we used a six month moving window. This method makes no attempt to capture the seasonality in the consumption data. All of the remaining methods that we considered aim to accommodate the seasonality in the consumption time series.

*3.4    Separate kernel density estimation for each period of the week (KD-W)*

To estimate the density for a given period of the week, this method applies KD estimation to historical observations that belong to the same period of the week. We refer to the method as KD-W. Specifically, to estimate the density for a period of week, denoted by $W$ ($W \in [1..s_2]$), we applied KD estimation to historical observations (lying within the moving window of $l$ periods) that fell on period $W$ of a past week. Furthermore, this method incorporates an exponential time decay parameter $\lambda$ ($0 < \lambda \leq 1$), which allows more emphasis to be placed on the more recent observations within the density estimation framework. This method is represented as:

$$\hat{f}(y) = \frac{\sum_{t \in [m,n] \mid t \bmod s_2 = W} \lambda^{\lfloor (n-t)/s_2 \rfloor} K_{h_y}(Y_t - y)}{\sum_{t \in [m,n] \mid t \bmod s_2 = W} \lambda^{\lfloor (n-t)/s_2 \rfloor}}, \qquad (3.5)$$

$h_y$ denotes the bandwidth for observations belonging to the same period of the week; $mod$ denotes the modulus operator; and $\lfloor \bullet \rfloor$ denotes the floor operator. Note that we have $s_1 = 48$ and $s_2 = 336$. The decay parameter imposes an exponential decay in uniform steps of size $s_2$. This ensures that observations belonging to the same week are given the same decay weight. A low decay parameter leads to a fast decay in the weights, thereby placing more emphasis on more recent observations. In addition to the bandwidth, we optimized the decay parameter using cross-validation. We included the seasonal discounting in all of the methods that we describe in the remainder of Section 3.

*3.5    CKD estimation conditional on the period of week (CKD-W)*

This method uses CKD to estimate the density by conditioning consumption on the period of the week. This method, which we term CKD-W, is represented as:

$$\hat{f}(y|w) = \frac{\sum_{t=m}^{n} \lambda^{\lfloor (n-t)/s_2 \rfloor} K_{h_{x\_week}}(W_t - w) K_{h_y}(Y_t - y)}{\sum_{t=m}^{n} \lambda^{\lfloor (n-t)/s_2 \rfloor} K_{h_{x\_week}}(W_t - w)}, \qquad (3.6)$$

where $h_{x\_week}$ is the bandwidth defined in the $x$-direction for the period of week, and $W_t = t \bmod s_2$. This method can be viewed as applying the CKD estimator to observations plotted against the period of week at which they occur, as in Fig. 4, which illustrates this for a 24-week period, for the SME considered in Fig. 2.

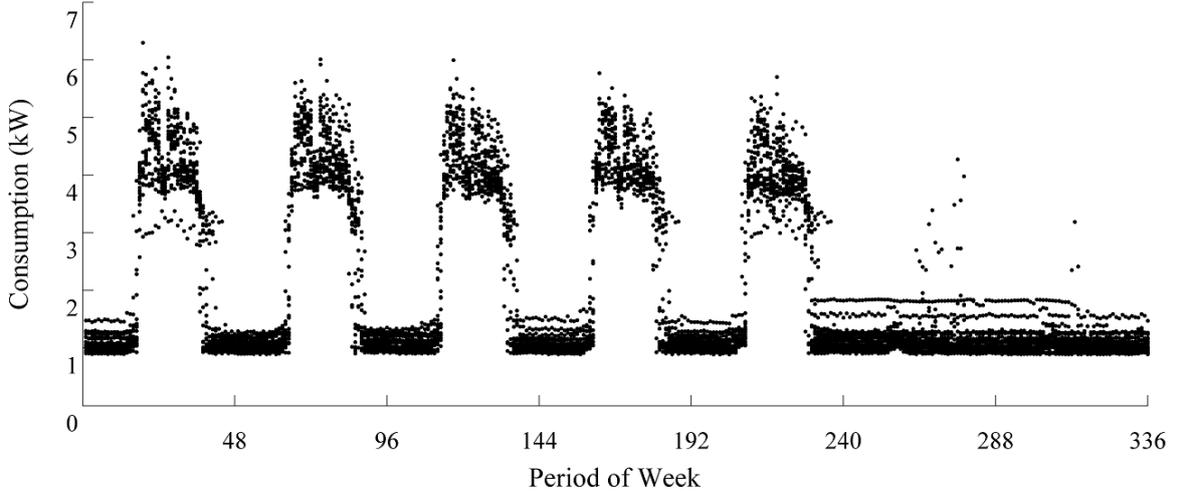

**Fig. 4.** SME consumption from 24 weeks of data, plotted against the period of the week, starting from the first period on a Monday.

The rationale of this method is that consumption observations are weighted in proportion to how close their period of occurrence is to the current period of the week. Note that, in this method, periods 1 and $s_2$ are treated as being a distance of 1 period apart, rather than $(s_2 - 1)$ periods apart. Hence, kernelling in the *x*-axis direction treats periods of the week $j$ and $k$ as being a distance apart equal to $\min(|j - k|, s_2 - |j - k|)$. This metric has been used within a CKD approach by Hyndman *et al*. (1996) for measuring the difference between two periods within a year.

*3.6  CKD estimation conditional on the period of week and period of day (CKD-WD)*

Time series of intraday electricity load, at the country level, have been observed to possess both intraweek seasonal cycles and, at least for the weekdays, repeating daily cycles. Parametric time series models have been proposed that aim to accommodate both of these forms of cycles (see Taylor, 2003; De Livera *et al*., 2011; Taylor and Snyder, 2012). With this same aim, we extended the CKD-W estimator of Section 3.5 so that it is conditional on both the period of the week, and the period of the day. This method, which we term CKD-WD, is represented as:

$$\hat{f}(y|w, d) = \frac{\sum_{t=m}^{n} \lambda^{\lfloor (n-t)/s_2 \rfloor} K_{h_{x\_week}}(W_t - w) \, K_{h_{x\_day}}(D_t - d) K_{h_y}(Y_t - y)}{\sum_{t=m}^{n} \lambda^{\lfloor (n-t)/s_2 \rfloor} K_{h_{x\_week}}(W_t - w) \, K_{h_{x\_day}}(D_t - d)}, \qquad (3.7)$$

where $h_{x\_day}$ is the bandwidth for conditioning on the period of the day $d$ ($d \in [1..s_1]$). We define $D_t$ as $t\ mod\ s_1$. The kernelling that conditions on the period of the day treats periods of the day $j$ and $k$ as being a distance apart equal to $\min(|j-k|, s_1 - |j-k|)$.

*3.7    KD estimation based on type of intraday cycle (KD-IC)*

This method treats a week as being composed of two different types of intraday cycle. Specifically, this method treats all five weekdays as having the same intraday cycle and it treats Saturdays and Sundays as having the same intraday cycle. The classification of different days of the week based on the similarity in their intraday cycle has been considered by Gould *et al.* (2008). Their parametric time series model has been termed intraday cycle (*IC*) *exponential smoothing*, and it is this that prompts us to label the method of this section as KD-IC. This method is represented as:

$$\hat{f}(y) = \frac{\sum_{t \in [m,n]} \lambda^{\lfloor (n-t)/s_2 \rfloor} K_{h_y}(Y_t - y)}{\sum_{t \in [m,n]} \lambda^{\lfloor (n-t)/s_2 \rfloor}}, \quad (3.8)$$

This method is similar to the KD-W method, the difference being that, instead of performing KD estimation on just the past observations belonging to the same period of the week, the KD-IC method performs KD estimation using past observations belonging to the same period on past days of the same day type.

*3.8    CKD estimation conditional on type of intraday cycle (CKD-IC)*

This method, which we refer to as CKD-IC, can be viewed as a synthesis of the CKD-W and KD-IC methods, such that: a) observations belonging to weekdays are modelled separately from weekends, and, b) two different bandwidths (in the $x$-direction) are used, with one for weekdays, and a distinct one for weekends. In Section 3.5, we explained that the CKD-W method can be viewed as applying the CKD estimator to the plot in Fig. 4, which shows observations for 24 past weeks plotted against each period of the week. For the CKD-IC method, the analogous plot is presented in Fig. 5 for the SME considered in Fig. 2. The plot is based on 24 weeks of data. In this plot, each period on a weekday has the 24 observations for the corresponding period on that particular weekday, plus the 24 observations for the corresponding period on each of the other weekdays, resulting in 120 observations plotted for each period on each weekday. Similarly, the observations for Saturday and Sunday are pooled so that 48 observations are plotted for each weekend period.

The inclusion of two different bandwidths in the $x$-direction allows observations to be smoothed differently for weekdays, compared to weekends.

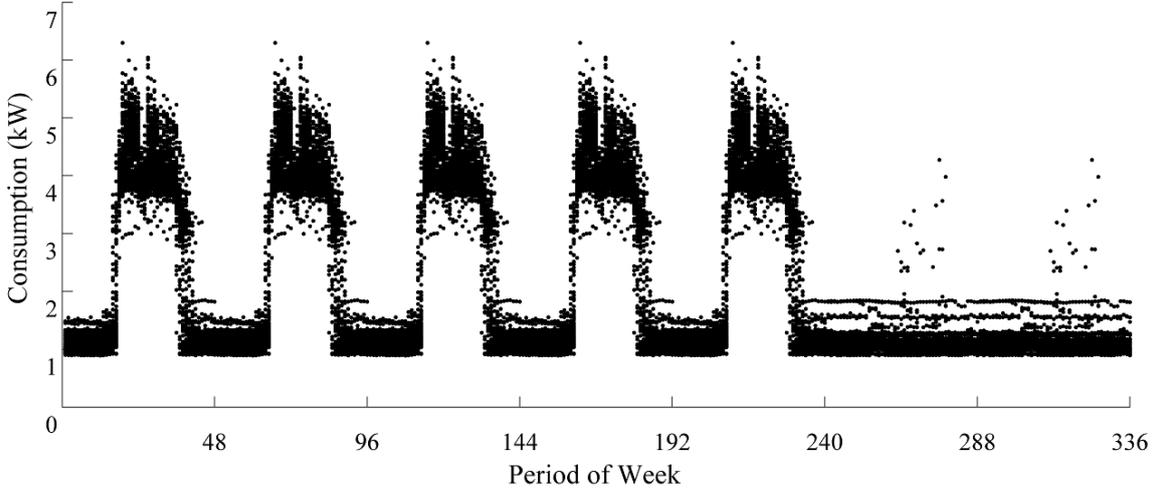

**Fig. 5.** SME consumption for 24 weeks, plotted against the period of the week, starting from the first period on a Monday. For each period, observations are plotted belonging to that period, as well as observations belonging to periods with the same intraday cycle.

*3.9    CKD estimation conditional on lagged consumption (CKD-Lag)*

Using this CKD method, the density is estimated by conditioning consumption for a given period, on the consumption observed during the same period from the previous week. This method, which we term CKD-Lag, is represented as:

$$\hat{f}(y|x) = \frac{\sum_{t=m}^{n} \lambda^{\lfloor(n-t)/s_2\rfloor} K_{h_{x\_lag}}(Y_{t-s_2} - x) K_{h_y}(Y_t - y)}{\sum_{t=m}^{n} \lambda^{\lfloor(n-t)/s_2\rfloor} K_{h_{x\_lag}}(Y_{t-s_2} - x)}, \quad (3.9)$$

where $h_{x\_lag}$ is the bandwidth defined in the $x$-direction for the lagged observations. This method can be viewed as the nonparametric analogue to seasonal autoregressive modelling. Hyndman *et al.* (1996) proposed a similar approach using CKD, whereby they modelled current daily temperature, conditional on the temperature observed the previous day.

*3.10    Parameter estimates*

As mentioned in Section 2, we estimated KD and CKD parameters for each of the tariff/stimulus categories using up to 10% of the consumers within each tariff/stimulus category. Note that the model parameters were optimized based on only in-sample density forecast performance, as quantified using the CRPS for the one-month cross-validation period. To summarise the optimised parameters, Table 2 presents the weighted average of the parameters, where the weights were chosen to be proportional to the number of individual residential consumers or SMEs in each category. (Note that averaging the parameters is reasonable, because we standardised consumption values to lie between 0 and 1.) The bandwidths in the $y$-direction are quite small, which is to be expected, as a substantial proportion of the observations are concentrated around zero. The values of the decay factor ($\lambda$) are close to one, which suggests that the weights decay slowly over time. For example, a discount factor of 0.942 corresponds to a half-life of approximately twelve weeks, and a discount factor of 0.977 corresponds to a half-life of approximately thirty weeks.

**Table 2**
Model parameters (bandwidths and decay parameter) for different methods averaged across residential consumers and SMEs.

|        | Parameters      | Residential consumers | SMEs  |
|--------|-----------------|-----------------------|-------|
| KD-U   | $h_y$           | 0.014                 | 0.061 |
| KD-W   | $h_y$           | 0.012                 | 0.038 |
|        | $\lambda$       | 0.942                 | 0.926 |
| CKD-W  | $h_{x\_week}$   | 0.909                 | 0.488 |
|        | $h_y$           | 0.014                 | 0.044 |
|        | $\lambda$       | 0.944                 | 0.917 |
| CKD-WD | $h_{x\_day}$    | 0.651                 | 0.354 |
|        | $h_{x\_week}$   | 0.553                 | 0.354 |
|        | $h_y$           | 0.013                 | 0.045 |
|        | $\lambda$       | 0.994                 | 0.925 |
| KD-IC  | $h_y$           | 0.014                 | 0.039 |
|        | $\lambda$       | 0.998                 | 0.917 |
| CKD-IC | $h_{x\_weekday}$| 0.704                 | 0.354 |
|        | $h_{x\_weekend}$| 0.825                 | 1.042 |
|        | $h_y$           | 0.015                 | 0.045 |
|        | $\lambda$       | 0.977                 | 0.938 |
| CKD-Lag| $h_{x\_lag}$    | 0.017                 | 0.045 |
|        | $h_y$           | 0.017                 | 0.045 |
|        | $\lambda$       | 0.958                 | 0.929 |

*3.11 Sophisticated Benchmark*

As a sophisticated benchmark, we produced point and density forecasts using the double seasonal Holt-Winters-Taylor (HWT) exponential smoothing method. This method allows the inclusion of an intraday cycle to be nested within an intraweek cycle. It has been used to model total national consumption (see, for example, Taylor, 2003; 2010). The HWT method requires the estimation of smoothing parameters for the level ($\lambda$), intraday seasonal index ($\delta$), intraweek seasonal index ($\omega$), and a parameter to adjust for residual autocorrelation ($\emptyset$). We estimate the HWT model parameters by maximizing a Gaussian likelihood, using the same training set and categorization scheme as used for the KD and CKD methods. For HWT, we generate density forecasts using Monte Carlo simulation with 10000 iterations. The weighted average of the parameter estimates for the residential consumers were: λ = 0.009, δ = 0.019, ω = 0.1203, and ø = 0.420, while for the SMEs, we obtained: λ = 0.014, δ = 0.050, ω = 0.139, and ø = 0.592.

## 4. Evaluating post-sample forecast accuracy

### 4.1 Evaluating density forecasts

In Fig. 6, for the residential consumer and SME considered in Figs. 1 and 2, we summarise the post-sample density forecasts for up to a week ahead. The forecast origin was the final half-hour on Sunday 15 August. Hence, horizons up to five days correspond to weekdays, and the longer horizons correspond to weekends.

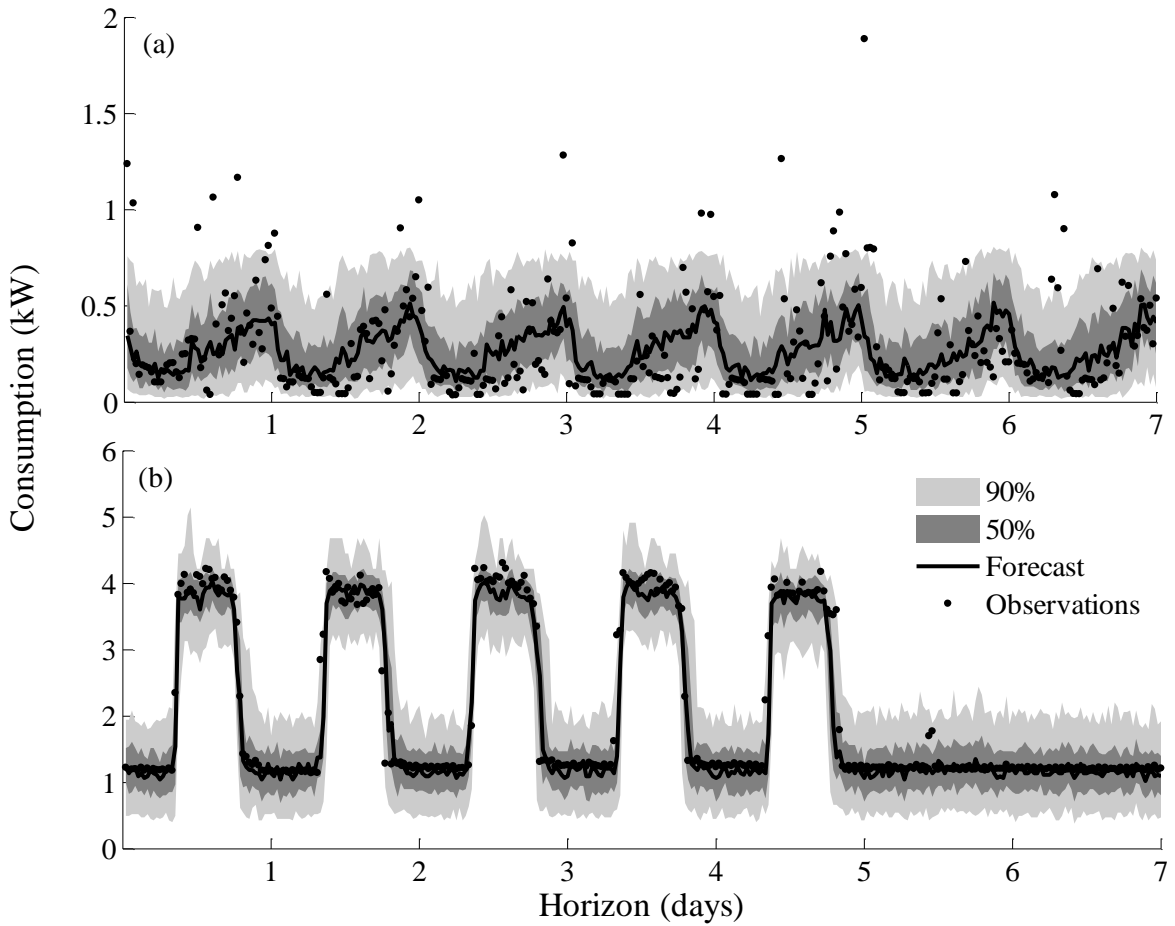

**Fig. 6.** Summaries of density forecasts for: a) a residential consumer, and, b) an SME.

In Fig. 6a, for the residential consumer, we used the KD-IC method to generate forecasts, as it was one of the most accurate methods across residential consumers. With similar motivation, we use the CKD-W method to generate density forecasts for the SME in Fig. 6b. Note that each point forecast was generated by taking the median of the corresponding density forecast. In both figures, it is encouraging to see the seasonal patterns that we identified in Fig. 3.

For density forecast evaluation, we use the CRPS, as it quantifies both *calibration* and *sharpness*, (see Gneiting *et al.*, 2007). It is defined as:

$$\text{CRPS}(\hat{F}_{t+h}, y_{t+h}) = \int_{-\infty}^{\infty} (\hat{F}_{t+h}(z) - \mathbf{1}(z \geq y_{t+h}))^2 dz, \quad (4.1)$$

where $y_{t+h}$ denotes the actual observation; $\hat{F}_{t+h}$ is the predicted cumulative distribution function at horizon $h$; and $\mathbf{1}$ is the indicator function. The CRPS is calculated for each period $t$ in the post-sample data.

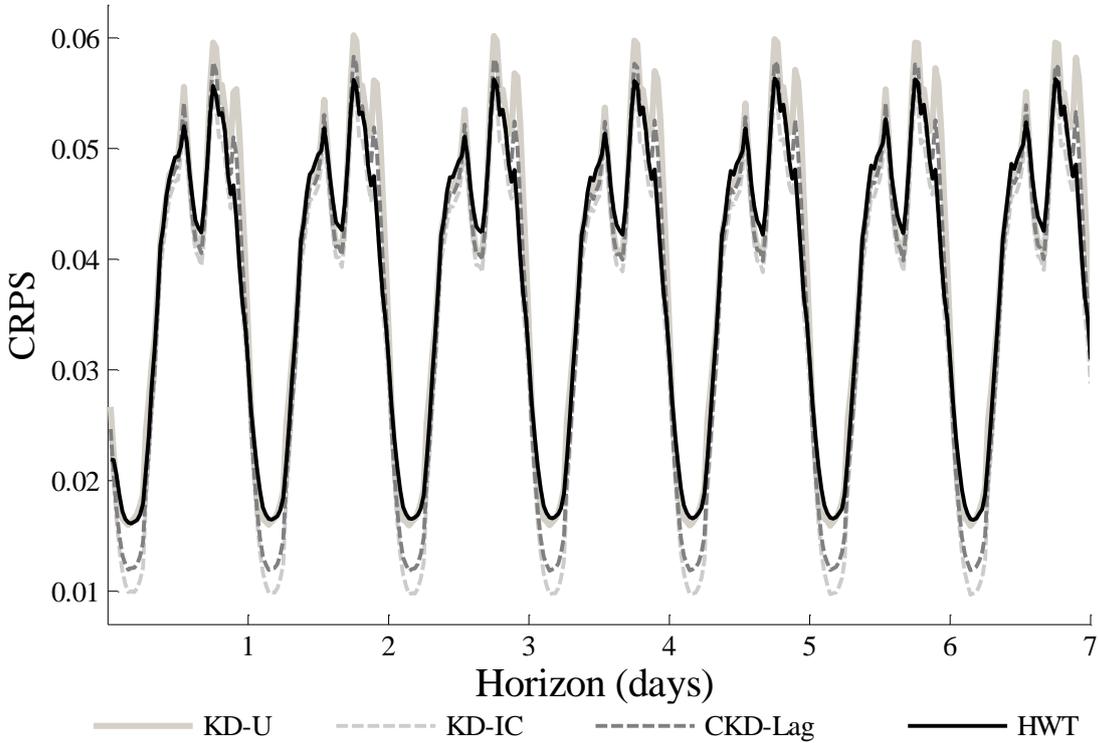

**Fig. 7.** Mean CRPS for density forecasting of the 800 residential consumers. Lower values are better.

Fig. 7 presents the post-sample CRPS results averaged across all 800 residential consumers, for lead times at half-hourly intervals, from one half-hour up to one week ahead. Averaging across our post-sample forecast accuracy measures was reasonable, because, for each series, we had standardized the consumption values to lie between 0 and 1. The forecast performance was similar for KD-W, CKD-W, CKD-WD, KD-IC and CKD-IC. For simplicity, of these five methods, we present results for only the KD-IC method, as it was one of the most accurate methods. As expected, the CRPS is high during periods of the week that witness a relatively large change in consumption. Among the CKD methods, the performance of CKD-Lag is the worst. It is interesting to see that the benchmark methods, namely KD-U and HWT, are not competitive with the other methods.

In Fig. 8, we present the post-sample CRPS results averaged across the 200 SMEs. The performance of the KD-IC and CKD-IC methods were very similar, and both methods were outperformed by the CKD-W and CKD-WD methods. Hence, for conciseness, we do not present results for the KD-IC and CKD-IC methods. The three best performing kernel-based methods, overall, were KD-W, CKD-W and CKD-WD. For simplicity, of these three methods, we present results for only the CKD-W method. Interestingly, the KD-U benchmark was substantially outperformed, which was not the case for the residential consumers, whereas the HWT benchmark, which did not perform very well for the residential consumers, is competitive with the best performing methods for the SMEs, as shown in Fig. 8. Our explanation for this result is that the residential consumption exhibits much greater volatility, making it relatively hard to model, whereas the consumption for the SMEs typically has prominent seasonality. Note that the KD-U benchmark method does not account for seasonality, whereas the HWT modelling framework is based on the assumption that the underlying data generating process exhibits prominent seasonality.

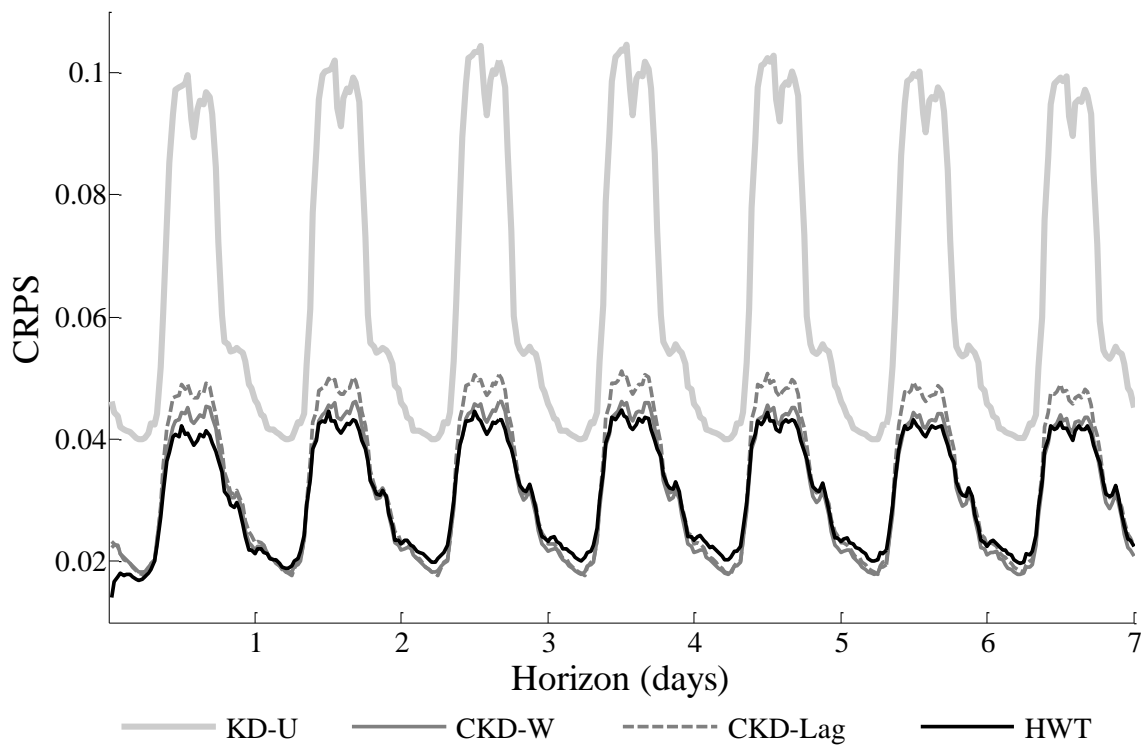

**Fig. 8.** Mean CRPS for density forecasting of the 200 SMEs. Lower values are better.

*4.2    Evaluating quantile forecasts*

We evaluate quantile forecasts using the unconditional coverage, which measures the percentage of observations that are lower than the corresponding $\theta$ quantile forecast. Ideally, this percentage should be $\theta$.

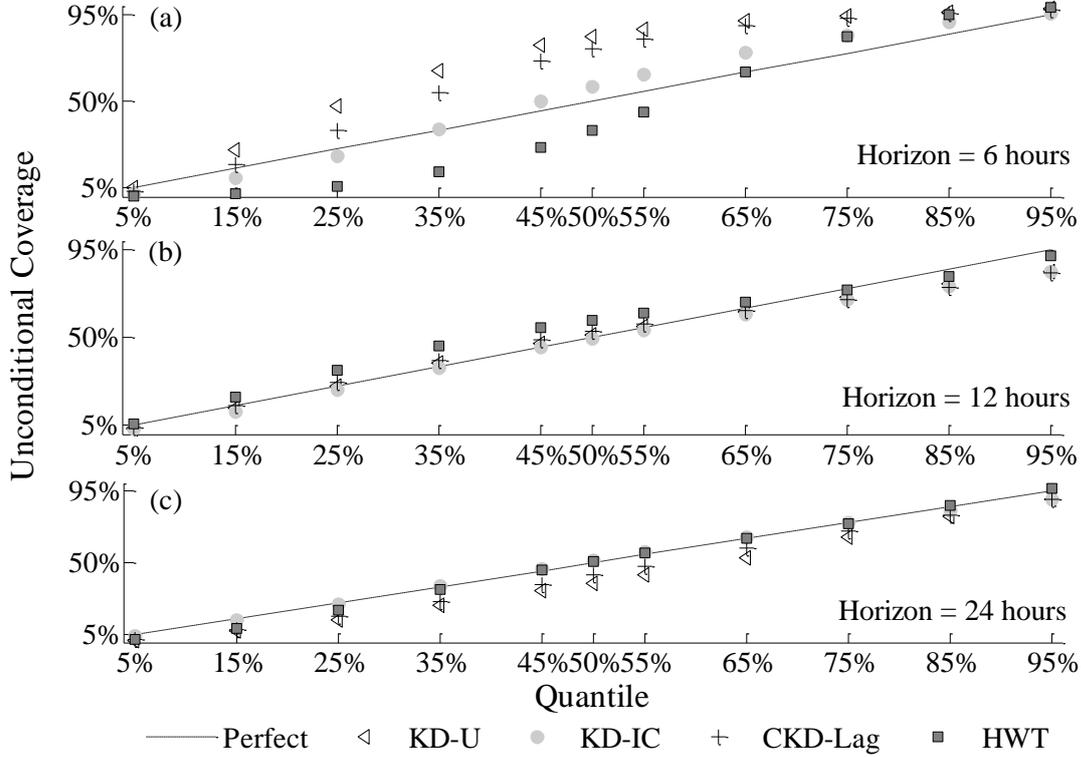

**Fig. 9.** Unconditional coverage for quantile forecasting for the 800 residential consumers for horizons: a) 6 hours, b) 12 hours, and, c) 24 hours. Values closer to the diagonal are better.

Fig. 9 presents the unconditional coverage, averaged across the 800 residential consumers, for the 5%, 15%, 25%, 35%, 45%, 50%, 55%, 65%, 75%, 85%, and 95% quantiles. The figure displays the results for three different lead times. Note that in Fig. 9, values closer to the diagonal line are better. In Fig. 9a, for 6 hour-ahead prediction, we can see that, for the KD-IC method, unconditional coverage is much closer to perfect coverage than for the KD-U method. Among the CKD based methods, CKD-Lag is the worst performing method. Fig. 9b shows that the unconditional coverage for twelve hour ahead forecasts is quite impressive for all the methods. For one-day ahead forecasts, Fig. 9c indicates that the unconditional coverage obtained using KD-IC and HWT is very good for all quantiles considered in this study. Fig. 10 presents the unconditional coverage averaged across the 200 SMEs. The CKD-W and HWT methods perform well, while KD-U and CKD-Lag show the weakest performance.

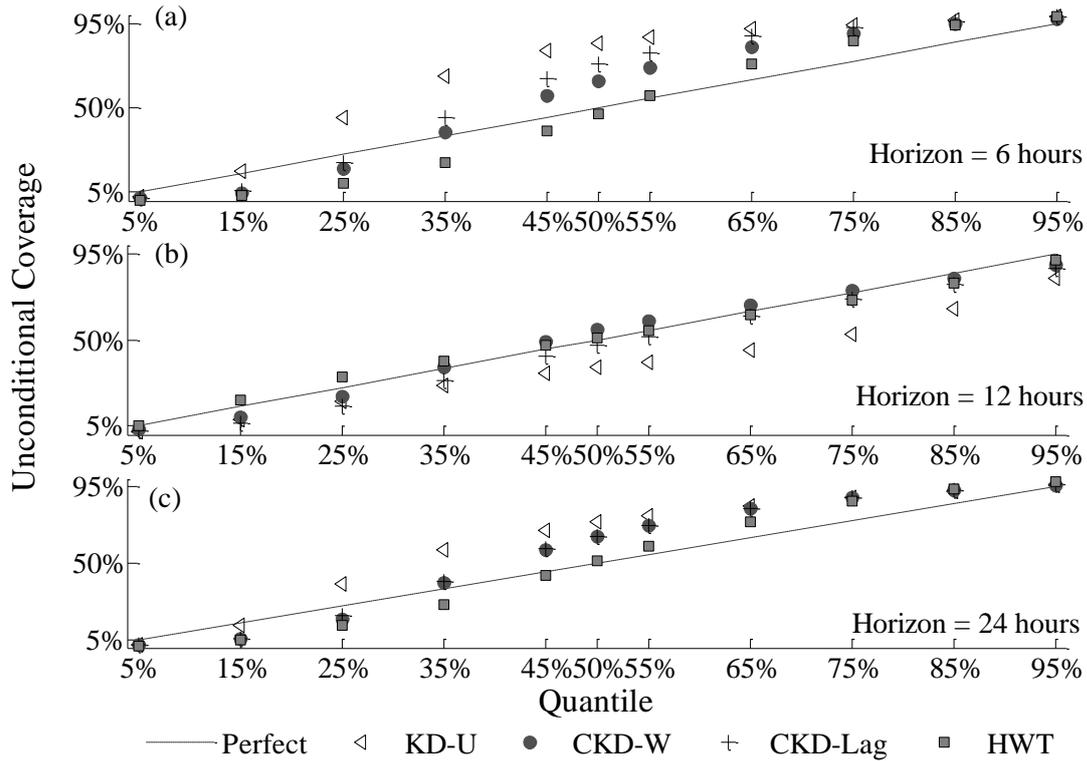

**Fig. 10.** Unconditional coverage for quantile forecasting for the 200 SMEs for horizons: a) 6 hours, b) 12 hours, and, c) 24 hours. Values closer to the diagonal are better.

### 4.3 Evaluating point forecasts

To evaluate point forecasts, we used the mean absolute error (MAE) and the root mean square error (RMSE). It has been pointed out by Gneiting (2011), that the median of the density forecast is the optimal forecast if the loss function is symmetric piecewise linear, whereas for a quadratic loss function, the mean is the optimal forecast. Hence, we issued the median of the density forecast as a point forecast, and evaluated it using the MAE. Similarly, we issued the mean of the density forecast as a point forecast, and used RMSE for evaluation. We found that the rankings of the different models across residential consumers and SMEs were very similar to the rankings obtained using the CRPS. Hence, for conciseness, we present the results of just the HWT sophisticated benchmark method and the best performing method for the residential consumers, which was KD-IC, and for the SMEs, which was CKD-W. The relative model rankings were very similar using the MAE and RMSE measures, and so we present results for only the MAE in this paper.

Fig. 11a presents the MAE for KD-IC and HWT, averaged across the 800 residential consumers. The KD-IC method is a little more accurate than HWT at nearly all horizons. Fig. 11b shows the MAE for CKD-W and HWT, averaged across the 200 SMEs. Interestingly, HWT is a little more accurate than CKD-W for lead times of less than one day, and also for lead times for which MAE was highest. As discussed previously, the enhanced performance of HWT across the SMEs can be attributed to the prominent seasonality in the consumption time series for the SMEs, as compared to the residential consumers.

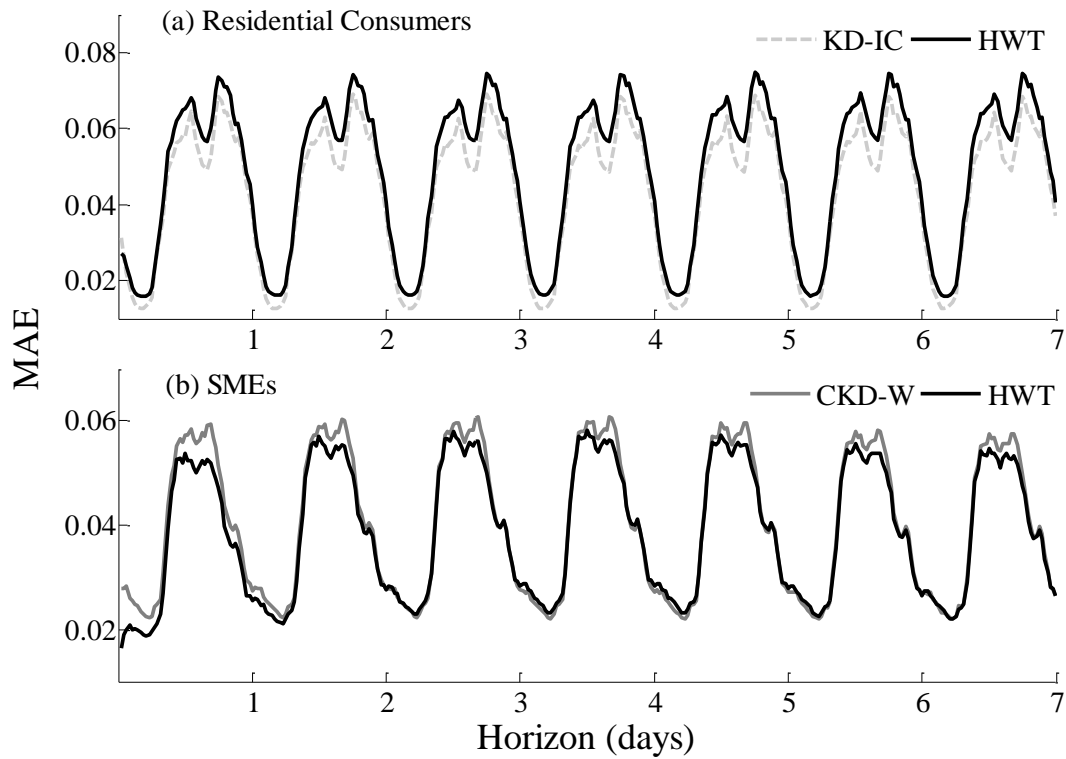

**Fig. 11.** Mean MAE for point forecasting of the: a) 800 residential consumers, and, b) 200 SMEs. Lower values are better.

In a practical setting, the efficacy of different methods discussed in this study would depend on their forecast performance and computational requirements. Thus, along with accuracy, it is crucial to gauge these methods in terms their scalability, and potential to be deployed for generating real-time online forecasts in an automated framework. The aspect of the kernel-based and exponential smoothing methods that is most computationally demanding is parameter estimation. However, we do not see this as an obstacle to implementation, because parameters would be estimated offline, and would be estimated, as in our study, for only a sample of the consumption series. Encouragingly, for a given set of parameters, the forecasts can be generated relatively swiftly. Specifically, the total elapsed CPU time to generate and evaluate forecasts for a typical residential consumer (using KD-IC) and an SME (using CKD-W) was approximately 11 seconds and 79 seconds, respectively. The methods were implemented in Matlab®, using a standard 64-bit (Intel i7, 2.8Ghz) machine.

# 5. Deriving prediction intervals for electricity cost

In this section, we illustrate how density forecasts of smart meter data could be used in the future to benefit consumers. We convert density estimates of consumption into density estimates of cost for different TOU tariffs. Comparison of different cost densities could help consumers identify and potentially select the tariff that is most suitable for their consumption patterns. For example, a risk-averse consumer might prefer a cost density with relatively low variance, even if it does not have the lowest mean. The importance for a decision maker to understand and interpret the uncertainty in a cost forecast has been emphasized by Kreye *et al*. (2012).

According to a report by DECC (2012), there are approximately 4.5 million fuel-poor households in the UK (2010 estimate), where a household is defined as fuel-poor if it needs to spend more than 10% of its income on fuel in order to maintain a satisfactory level of temperature throughout the home. It has been reported that 90% of the households in the lowest income group use prepayment as a mode for paying bills, as opposed to using credit payment (see, Waddams Price, 2002). Darby (2012) points out that although unit energy prices are higher under prepayment schemes, customers opt for this mode of payment because it gives them greater control over their consumption. It is thus crucial, especially for prepayment consumers, to be able to gauge the potential energy costs that would be incurred in the future. In this regard, future cost estimates can potentially help the consumers: a) select between credit and prepayment schemes, b) identify and select the most suitable tariff, and, c) adjust their energy usage to try to ensure that the cost does not exceed the stipulated budget.

To generate an estimate of the probability density function for cost, for a given period, we first sample realizations from the estimated consumption density. We then multiply the sampled consumption values (in kW), with the electricity cost (cents per kWh) associated with the period under consideration (for a given TOU tariff). The histogram of the resultant cost values provides an estimate of the density function for cost.

In Fig. 12, we summarise the cost density forecasts for the same residential consumer and the same forecast origin considered in Fig. 6a. As in that figure, in the forecast period of Fig. 12, the first five days correspond to weekdays, followed by the weekend. Each point forecast was produced as the median of the corresponding predicted density. In Fig. 12, we compute prediction intervals for cost for the residential consumer using Tariff B, as this consumer was allocated Tariff B by the CER. Note that we use the KD-IC method to compute prediction intervals for cost in Fig. 12, as it was one of the most accurate methods across the residential consumers. The cost associated with a given tariff is highest during peak-times (5pm-7pm) on weekdays (as shown in Table 1). It is for this reason, that there are five distinctive peaks observed during peak-times on weekdays, as shown in Fig. 12.

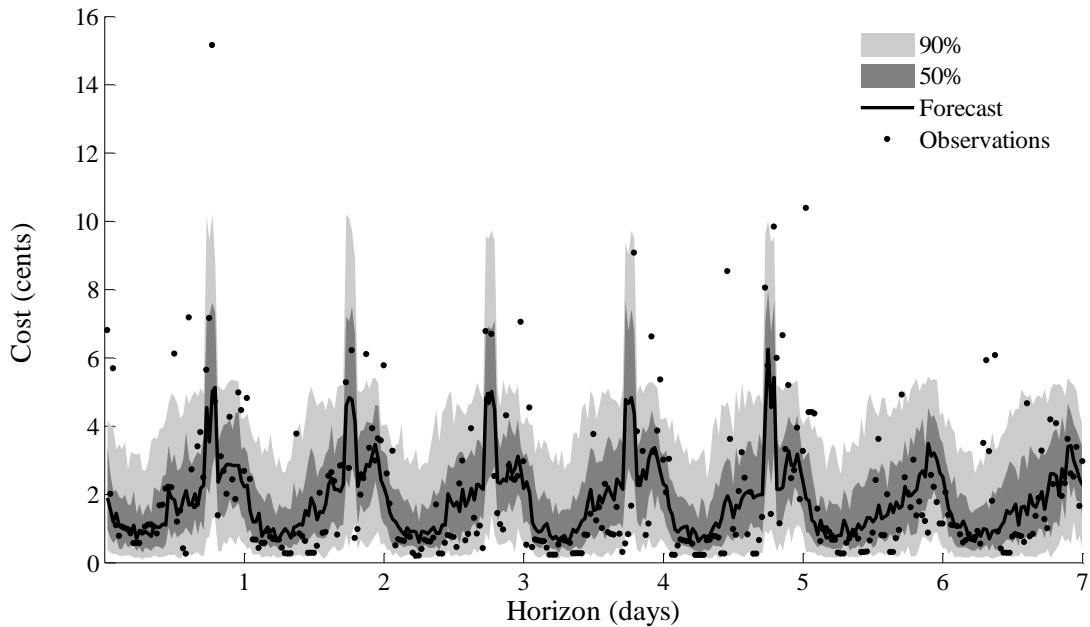

**Fig. 12.** For a residential consumer, summaries of cost density forecasts computed using the KD-IC method.

Of the 800 residential consumers in our dataset, 220 belonged to the control group. These control group consumers were not allocated TOU tariffs by the CER (CER, 2011a). Instead, the control group was billed on their normal electricity supplier tariff. We did not have information regarding the tariff applicable for each control group consumer. Hence, in our empirical work, we included only the 580 residential consumers that were not in the control group. We used data from only the residential consumers for this analysis, because we did not have information regarding the specific TOU tariff allocated to each SME by the CER.

In addition to the five tariffs described in Table 1, we investigated a non-TOU tariff, which we refer to as Tariff E. For this tariff, we set the cost to be 13.25 cents per kWh, for each period of the day. We devised this cost by computing the weighted mean of the cost associated with different periods of the day, using Tariff D, whereby the weights were proportional to the number of half-hours occurring during the day, peak and night periods. Hence, for a given density forecast of consumption, we derived six density estimates of cost, each one corresponding to a different tariff

We then selected a tariff by comparing the six cost densities. To compare cost densities, we considered the following three alternative criteria: (1) lowest mean (expectation), (2) lowest 75% quantile, and, (3) lowest 95% quantile. Selecting a tariff using the mean criterion can be seen as a *risk neutral* strategy, while the 75% and 95% quantile criteria can be viewed as *risk averse* and *very risk averse* strategies, respectively.

We investigated whether it is more cost effective for the consumers to switch between different tariffs or to stick with the tariff allocated to each consumer by the CER. To compute the cost associated with switching, we selected the optimum tariff for the coming week (each Sunday midnight) based on the comparison of different predicted cost densities. We then computed the cost associated with the selected tariff for the whole week under consideration, using the actual consumption observations. For each consumer, we repeated this procedure for all four weeks in the post-sample period to compute the total cost associated with switching.

Table 3 presents the percentage of consumers who were able to achieve lower cost by switching tariff; the percentage incurring lower costs by sticking with the allocated tariff; and the percentage for which there was no difference between switching and sticking with the allocated tariff. From the table, we see that switching resulted in cost savings for the great majority of consumers, regardless of which of the three different criteria was used. Interestingly, the average cost saving from switching was noticeably more when using the 75% and 95% quantiles as the basis for switching, rather than the mean of the predicted cost densities. This illustrates the value in generating density forecasts for consumption recorded by smart meters.

**Table 3**
Percentage of consumers who achieved lower costs by: a) switching tariffs, b) using the allocated tariff, and, c) who found no difference between the two schemes. The table also presents the average savings achieved via switching tariffs (in cents per consumer) averaged across all 580 residential consumers.

|  | Mean (*Risk Neutral*) | 75% Quantile (*Risk Averse*) | 95% Quantile (*Very Risk Averse*) |
|---|---|---|---|
| Switching Tariff | 74.1% | 90.2% | 92.1% |
| Allocated Tariff | 10.0% | 4.3% | 3.8% |
| No Difference | 15.9% | 5.5% | 4.1% |
| Average Saving from Switching | 168 | 202 | 211 |

# 6. Summary and concluding remarks

In this study, we modelled the density of electricity smart meter data using different implementations of the KD and CKD estimators. The methods were aimed at accommodating the seasonality in consumption, along with the underlying variability. We used a decay parameter in the modelling framework to place more emphasis on the more recent observations. The evaluation of post-sample density and quantile forecasts showed that the methods considered in this study convincingly outperformed the unconditional KD estimator. Encouragingly, the employed methods were able to accommodate the weekly seasonality that can be present in consumption. We also implemented the HWT exponential smoothing method, and found that point and density forecasts from this method were particularly competitive for the SME series. This can be attributed to the relatively strong seasonal patterns typically present in the SME series. Furthermore, we derived density forecasts for electricity cost using the density forecasts of electricity consumption, for six different tariffs. Using three different tariff selection criteria, our empirical study showed that switching between tariffs results in considerable cost savings overall, compared to the case when consumers are allocated a single tariff for all periods.

A potentially useful line of future work would be to accommodate holiday effects while estimating the density of electricity consumption, where each special day is modelled separately. Another potential area of research would be to investigate the incorporation of different clustering techniques, such as those considered by Dai and Kuosmanen (2014), to identify consumers with similar electricity consumption patterns. This information could potentially be useful for scaling the modelling framework to a large number of consumers, without incurring huge computational costs.

**Acknowledgement**

We are grateful to the CER for making the anonymized smart meter data available. We would also like to thank two anonymous referees for providing valuable comments.